\documentclass[aps,prd,twocolumn,superscriptaddress,amsmath,widetext,amssymbepsfig,showpacs]{revtex4-1}
\usepackage{graphicx}
\usepackage{dcolumn}
\usepackage{bm}
\usepackage{natbib}

\newcommand{\be}{\begin{equation}}
\newcommand{\ee}{\end{equation}}
\newcommand{\bea}{\begin{eqnarray}}
\newcommand{\eea}{\end{eqnarray}}

\newcommand{\LCDM}{$ \Lambda $CDM}

\newcommand{\threej}[6]{{\left( \begin{array}{ccc} #1 & #2 & #3 \\ #4 &
   #5 & #6 \end{array} \right)}}

\urldef\mgcamb\url{http://www.sfu.ca/~aha25/MGCAMB.html}

\begin{document}
\title{Parametrised modified gravity and the CMB Bispectrum}

\author {Eleonora Di Valentino}
\affiliation{Physics Department and INFN, Universit\`a di Roma ``La Sapienza'', Ple Aldo Moro 2, 00185, Rome, Italy}

\author {Alessandro Melchiorri}
\affiliation{Physics Department and INFN, Universit\`a di Roma ``La Sapienza'', Ple Aldo Moro 2, 00185, Rome, Italy}

\author {Valentina Salvatelli}
\affiliation{Physics Department and INFN, Universit\`a di Roma ``La Sapienza'', Ple Aldo Moro 2, 00185, Rome, Italy}

\author {Alessandra Silvestri}
\affiliation {Department of Physics, MIT, Cambridge, MA 02139 USA}

\begin {abstract}
We forecast the constraints on modified theories of gravity from the cosmic microwave
background (CMB) anisotropies bispectrum that arises from correlations
between lensing and the Integrated Sachs-Wolfe effect. In models of modified gravity the evolution of the metric potentials is generally altered and the contribution to the CMB bispectrum signal can differ significantly from the one expected in the standard cosmological model.
We adopt a parametrised approach and focus on three different classes of models: Linder's growth index, Chameleon-type models and $f(R)$ theories.  We show that the constraints on the parameters of the models will significantly improve with  future CMB bispectrum measurements.
\end {abstract}


\maketitle

\section {Introduction} \label {sec:intro}

Cosmic acceleration is one of the major challenges faced by modern cosmology and understanding the very nature of what is sourcing it is the main focus of upcoming and future cosmological experiments.  
Several  approaches to the phenomenon of cosmic acceleration have been proposed in the literature, including modifications of the laws of
gravity on large scales in order to allow for self-accelerating
solutions in matter-only universes. Well-known examples of modified theories of gravity are 
 $f(R)$ models~\cite{Starobinsky:1980te, Capozziello:2003tk, Carroll:2003wy, Starobinsky:2007hu, Nojiri:2008nk},  or the more general scalar--tensor theories~\cite{EspositoFarese:2000ij, Bartolo:1999sq, Elizalde:2004mq, Chiba:2003ir},
the Dvali-Gabadadze-Porrati (DGP) model~\cite{Dvali:2000hr, Deffayet:2000uy}, and its further extensions such as Degravitation~\cite{Dvali:2007kt}. In the past years several authors have analyzed constraints on modified gravity, or more generally departures from the cosmological standard model, both using current datasets as well as doing forecasts for future surveys~\cite{Song:2006ej,Bean:2006up,Pogosian:2007sw,Tsujikawa:2007xu,Zhao:2008bn,Lue:2003ky,Koyama:2005kd, Song:2006jk, Song:2007wd, Cardoso:2007xc, Giannantonio:2008qr,caldwell2007,daniel2008,giannantonio2009,martinelli2010,
daniel2010} 
 
In this paper we focus on certain classes of modified gravity and adopt a parametrised approach to forecast the 
constraints achievable from measurements of the cosmic microwave background (CMB) bispectrum from future experiments.

Future high-resolution CMB maps will have the ability of detecting 
higher-order correlations in the temperature distribution at high significance (see e.g. \cite{komatsuNG}
and references therein).
While the CMB anisotropy distribution is generally expected to be Gaussian to high accuracy, 
small non-Gaussianities  could be produced in the early universe, during
inflation (commonly referred to as primordial non-Gaussianities, see e.g. \cite{bartolo}) as well as be sourced, 
at a much later epoch, by the interaction of CMB photons with the local universe. 
For instance, the lensing of CMB photons by dark matter structure 
produces a clear non-Gaussian signal in the CMB trispectrum (the Fourier transform of the four-point correlation function), which can be 
 used to constrain the amplitude of the lensing potential. Such signal, already
discovered by the recent ACT~\cite{act} and SPT~\cite{spt} experiments helps
in further constraining cosmological models.

In this paper we study the implications of another non-Gaussian signal expected in the CMB, {\it i.e.} the one
arising from cross-correlations between lensing and the Integrated Sachs-Wolfe effect (ISW), 
which affects the CMB bispectrum, {\it i.e.} the three-point correlation function.

The signature of the lensing-ISW (L-ISW hereafter) correlations in the CMB bispectrum
 has already been discussed by several authors (see e.g.~\cite{mollerach,mollerach2,spergold,sperver,gold,giovi,smithzal,serra2008,mangilli2009,hanson,lewis2011})

Assuming the standard cosmological scenario, $\Lambda$CDM, the L-ISW bispectrum should be
detected at the significance level of $4 - 5$ standard deviations. In~\cite{spergold,gold,giovi} the possibility of constraining cosmological parameters
through this detection has been considered; in particular it has been shown
that an accurate measurement of the L-ISW will help in constraining the equation of
state and the fractional density of dark energy. Here we shall analyze the potential of 
the L-ISW signal to constrain modified theories of gravity. In the latter,  the 
evolution of the metric potentials can generally differ significantly from the $\Lambda$CDM prediction, therefore it is natural to expect that the L-ISW bispectrum would provide valuable constraints on these theories.

The paper is organized as follows. In Section~\ref {sec:theories} we present the set of modified gravity models considered for our analysis while in Sec.~\ref{sec:bispectrum} we review the L-ISW bispectrum. In Section~\ref{sec:analysis}
we describe the analysis method and in Sec.~\ref {sec:constraints} we present our results. We conclude in Section \ref {sec:concl}.

\section {Parametrised Modified Gravity} \label{sec:theories}
Many models of modified gravity have been proposed as alternatives
to $\Lambda$CDM, and analysing them one by one is impractical. The idea behind parametrised versions of modified gravity is exactly that  of encompassing several models into a single framework. The parametrisations that we consider for our analysis cover a fairly large sample of theories and allow us to draw quite general conclusions about the constraining power of the data considered. 

In our analysis we fix the background to that of the
$\Lambda$CDM model of cosmology. The latter is currently in very good agreement with all observables constraining the expansion history, and many models of modified gravity can mimic it while introducing significant modifications at the level of perturbations. Therefore, fixing the background to \LCDM, allows us to isolate the effects of departures at the level of growth of structure, where we expect the most significant deviations.

\subsection{Linder model}
In~\cite{linder2007} Linder introduced a simple parametrisation of the growth of density perturbations in the linear regime, via a single parameter, the growth index $\gamma$ (which we will denote with $\gamma_L$), defined through

\begin{equation}\label{gammaL}
g(a)=e^{\int_0^a d ln{a}[\Omega_m(a)^{\gamma_{\rm L}}-1]}
\end{equation}
The idea is that of capturing independently the information from the expansion and the growth history, respectively in $\Omega_m$ and $\gamma_L$.  Since in our analysis we fix the background to \LCDM, $\Omega_m(a)$ is determined by that and the only parameter of interest will be $\gamma_L$.

In the cosmological concordance model, \LCDM, as well as in vanilla-type dark energy models, $\gamma_L$ is to good approximation constant and equal to $\gamma_L\approx 6/11$. While it can generally be a function of time and scale, in several models of modified gravity it can still be approximated by a number, which often differs significantly from the $\Lambda$CDM value.  For instance in the braneworld gravity of the DGP model, $\gamma_L\approx 0.68$ to good approximation over the whole history~\cite{linder2007}. Things are more complicated for scalar-tensor models where often the time- and scale-dependence of $\gamma_L$ cannot be neglected. However, $\gamma_L$ remains a powerful \emph{trigger} parameter, since any deviation of it from $\approx 6/11$ would indicate a breakdown of the cosmological concordance model. 

As a starting point for our analysis, we assume $\gamma_L\approx const.$ and forecast constraints on this simple one parameter model.

\subsection {Chameleon-type models}
Chameleon-type theories correspond to gravity plus a scalar degree of freedom which is conformally coupled to matter fields, and has therefore a profile and a mass which depends on the local density of matter. The common action for such theories is
\begin{eqnarray}\label{action_gen}
S &=& \int  d^4 x \sqrt {-g} \left[ \frac{M_P^2}{2}R-\frac{1}{2}g^{\mu\nu}\left(\nabla_{\mu}\phi\right)\left(\nabla_{\nu}\phi\right)-V(\phi)\right]\nonumber\\
&&+S_i\left(\chi_i,e^{-\alpha_i(\phi)/M_P}g_{\mu\nu}\right)
\end{eqnarray}
 where $\phi$ is the scalar d.o.f., $\chi_i$ is the $i^{\textrm{th}}$ matter field and $\alpha_i(\phi)$ is the coupling of $\chi_i$ to $\phi$. We will limit ourselves to cases in which the coupling is a linear function of the scalar field, {\it i.e.} $\alpha_i(\phi)\propto\beta_i\phi$. A well known example of the latter are $f(R)$ theories as we discuss in Sec.~\ref{subsec:fR}

The free parameters of these theories are the mass scale of the scalar field and the couplings $\beta_i$. Since we consider constraints from late time cosmology, we are interested only in the  coupling to dark matter, and therefore drop the index $i$. 

While the modifications enter through the coupling of the scalar field to matter, and therefore change the energy-momentum conservation equations, it is possible to keep the latter unchanged and effectively absorb the modifications of the evolution of perturbations in the Poisson and anisotropy equation. The latter are commonly parametrised with two functions $\mu$ and $\gamma$, as follows
 \bea\label{mu}
 &&k^2\Psi=-\frac{a^2}{2M_P^2}\mu(a,k)\rho\Delta \ ,\\
 \label{gamma}
&&\frac{\Phi}{\Psi}=\gamma(a,k) \,,
 \eea
where $\rho\Delta\equiv\rho\delta+3\frac{aH}{k}(\rho+P)v$  is the comoving density perturbation. Furthermore, for Chameleon-type theories $\mu$ and $\gamma$ are well represented by the parametrisation introduced in~\cite{Bertschinger:2008zb} 
\bea\label{par_mu_gamma}
&&\mu(a,k)=\frac{1+\beta_1\lambda_1^2\,k^2a^s}{1+\lambda_1^2\,k^2a^s}\,,\nonumber\\
&&\gamma(a,k)=\frac{1+\beta_2\lambda_2^2\,k^2a^s}{1+\lambda_2^2\,k^2a^s}\,,
\eea
where the parameters $\beta_i$ can be thought of as dimensionless couplings, $\lambda_i$ as dimensional
length-scales  and $s$ is determined by the time evolution of the characteristic lengthscale of the theory, {\it i.e.} the mass of the scalar d.o.f.
As shown in~\cite{Zhao:2008bn}, in the case of Chameleon-type theories the parameters 
$\{\beta_i,\lambda_i^2\}$ are related in the following way
\be\label{ST_relation}
\beta_1=\frac{\lambda_2^2}{\lambda_1^2}=2-\beta_2\frac{\lambda_2^2}{\lambda_1^2}
\ee
and $1\lesssim s\lesssim4$, so that effectively the degrees of freedom are one coupling and a time-evolving lengthscale.

\subsection{$f(R)$ theories}\label{subsec:fR}.

As it becomes clear in the Einstein frame, $f(R)$ theories are a subclass of the models described by action~(\ref{action_gen}), corresponding to a universal fixed coupling $\alpha_i=\sqrt{2/3}\,\phi$. Therefore they can also be described by the parametrisation in (\ref{par_mu_gamma}). It can be easily seen that the fixed coupling $\alpha_i=\sqrt{2/3}\,\phi$ gives $\beta_1=4/3$ and $\beta_2=1/2$. Viable $f(R)$ models that closely mimic $\Lambda$CDM  have $s\sim 4$~\cite{Zhao:2008bn}, therefore the number of free parameters in Eqs.~(\ref{par_mu_gamma}) can be effectively reduced to one lengthscale (using (\ref{ST_relation})  e.g. the lengthscale $\lambda_1$. 

The latter is directly related to the mass scale of the scalar degree of freedom introduced by these theories and represented by $f_R\equiv df/dR$, known as the \emph{scalaron}. Specifically, $\lambda_1$ sets the inverse mass scale of the scalaron today,  {\it i.e.} $\lambda_1=1/m_{f_R}^0$. The results in the literature are usually presented in terms of a parameter $B_0$ which is related to $\lambda_1$ as follows~\cite{Yong-Hu-Sawicki}

\begin{equation}\label{B0}
B_0=\frac{2H_0^2\lambda_1^2}{c^2}
\end{equation}

Studying this particular subclass is interesting because some models belonging to this category have been shown to be  cosmologically viable and pass local tests of gravity~\cite{Hu-Sawicki}.

\section{The lensing-ISW bispectrum}\label{sec:bispectrum}
As it is well known, if the expansion of the universe is not
matter dominated ({\it i.e.} $\Omega_m\neq1$), the time variation of the
gravitational potential provides an additional source of CMB
anisotropies; restricting to the linear regime, this effect is known as the Integrated Sachs Wolfe (ISW) effect~\cite{Sachs:1967er}, given by:

\begin{equation}
\frac{\delta T}{T}({\bf n})|_{ISW} = \int d\chi\,  
\left(\Psi+\Phi\right)_{,\tau}({\bf n},\chi).
\label{eqn:isw}
\end{equation} 

\noindent where $\tau$ is conformal time. This adds a secondary anisotropy to the primordial signal and a significant contribution from it is expected at late times, when the Universe starts accelerating. 

Furthermore, the weak gravitational lensing by matter density fluctuations between us and the last scattering surface shifts the observed direction of photons. In practice, 
 the temperature anisotropy measured by an observer in the direction
${\bf n}$ is actually the anisotropy in the direction
$({\bf n} + \nabla\phi({\bf n}))$, {\it i.e.}

\begin{equation}\label{lensed_temp}
\delta \tilde{T}({\bf n}) =\delta T({\bf n} + \nabla \phi({\bf n}))
\end{equation}

\noindent where $\delta \tilde T({\bf n})$ is the lensed anisotropy and $\delta T({\bf n})$ is the unlensed one (primordial plus ISW). The deflection angle is written in terms of the \emph{lensing potential} $\phi$, which is a weighted integral of the Weyl potential $\Phi+\Psi$ over the line of sight:

\begin{equation}
\phi ({\bf n}) = \int_0^{\chi_*} d\chi \, g(\chi) 
\left[ (\Psi+\Phi)({\bf n},\chi)\right].
\label{eqn:lens}
\end{equation}
Here $\chi$ is the comoving distance from the observer and $g(\chi) =
(\chi_*-\chi)/(\chi\chi_*)$ with $\chi_*$ the distance at the last scattering
surface.

The Weyl potential enters both in the ISW and lensing kernel, therefore the two effects are correlated and they contribute a non-zero third order statistics in the CMB, {\it i.e.} the L-ISW bispectrum. 
\begin {figure}[t]
\includegraphics[width=0.65\linewidth, angle=-90]{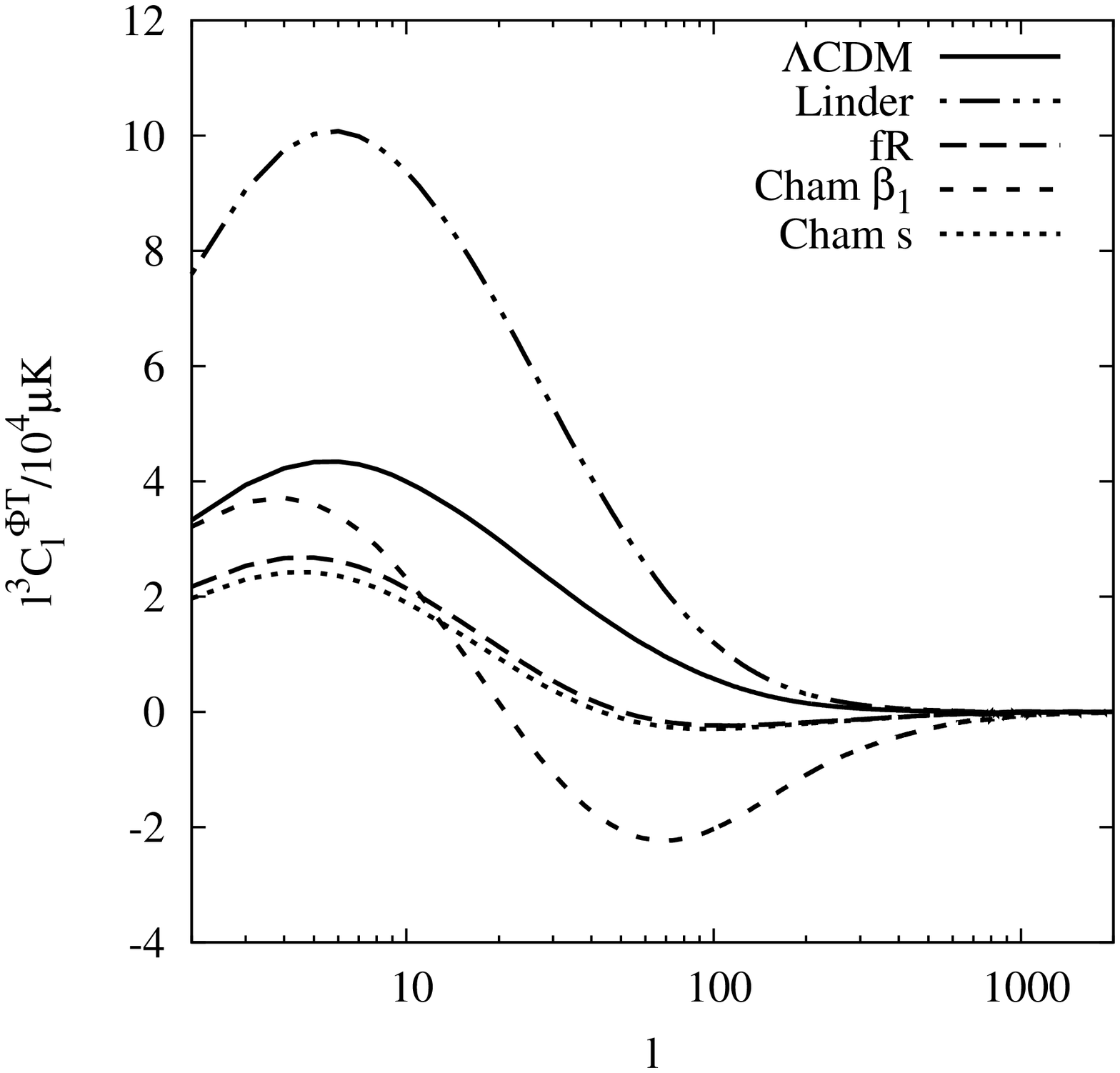}
\includegraphics[width=0.65\linewidth, angle=-90]{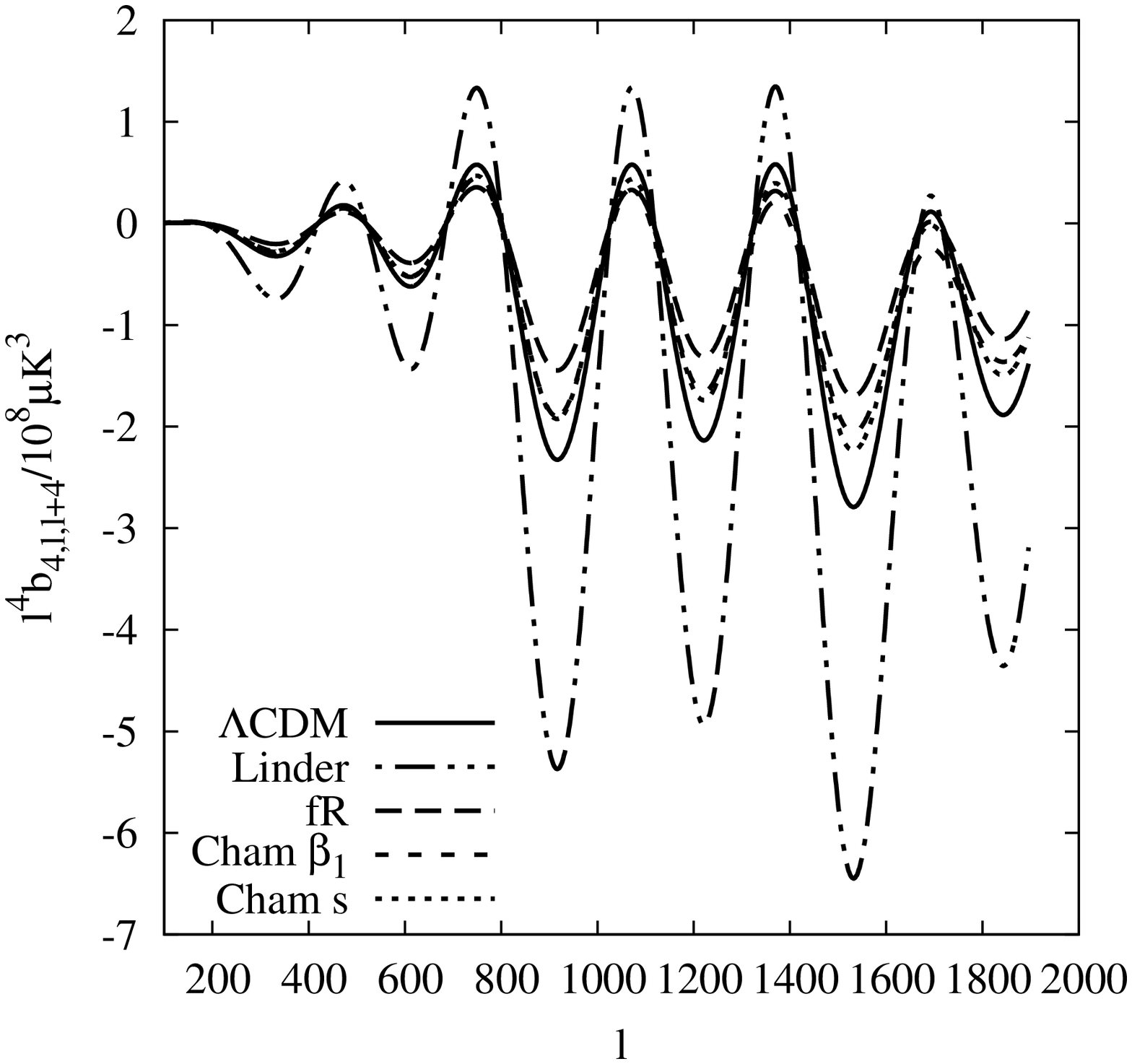}
\caption{Dependence of the cross temperature-lensing $C_\ell^{T\phi}$ angular spectra (top panel) and of the
reduced bispectrum (bottom panel) on modified gravity parameters for the different models considered in the analysis. The solid curves correspond to $\Lambda$CDM, the dotted and dashed curves to the Linder parametrisation with $\gamma_L=0.645$, the long-dashed curves to an f(R) model with $B_0=0.42$, the short-dashed curves to a Chameleon model 
(Cham${\beta_1}$) with $\beta_1=1.3$, $B_0=0.50$ $s=2.0$, while the dotted line to a Chameleon model
(Cham ${s}$) with  $\beta_1=1.3$, $B_0=0.50$, $s=3.3$.}
\label{fig:models}
\end{figure}

As usual, it is convenient to consider an expansion in spherical harmonics of the temperature field:

\begin{equation}
\frac{\delta T}{T}({\bf n}) = \sum_{\ell=2}^\infty \sum_{m=-\ell}^\ell a_{\ell m} Y_{\ell m}({\bf n})
\end{equation}

\noindent as well as of the lensing potential, $\phi({\bf n})=\sum_{\ell,m}\phi_{\ell m}Y_{\ell m}({\bf n})$. Taylor expanding Eq.~(\ref{lensed_temp}) in the lensing potential, and applying the above harmonic expansions, one obtains the following relation between the lensed and unlensed multipole coefficients, (to first order in in the lensing multipoles):

\begin{equation}\label{lensed_multipoles}
{\tilde a}_{\ell_1 m_1} \approx a_{\ell_1 m_1} + \sum_{\ell_2m_2\ell m} f_{\ell_1\ell_2\ell}\, a_{\ell_2m_2}^* \phi_{\ell m}^*
\left(
  \begin{array}{ccc}
    \ell_1 & \ell_2 & \ell \\
      m_1 & m_2 & m
  \end{array}
\right) 
\end{equation}
\noindent where the coefficient $ f_{\ell_1\ell_2\ell}$ is given by

\be
f_{\ell_1\ell_2\ell} = 
   \left( \frac{-\ell_1(\ell_1+1) + \ell_2(\ell_2+1) + \ell(\ell+1)}{2} \right)\,\Upsilon_{\ell_1\ell_2\ell}
\ee
   
\noindent with 

\be\label{gaunt_reduced}
\Upsilon_{\ell_1\ell_2\ell_3}\equiv\sqrt{\frac{(2\ell_1+1)(2\ell_2+1)(2\ell_3+1)}{4\pi}}\left(\begin{array}{ccc}\ell_1&\ell_2&\ell_3\\0&0&0\end{array}\right) \ .
\ee

Then, the angle averaged CMB bispectrum generated by the lensing-ISW correlation is given by

\begin{eqnarray}\label{angle_averaged_B}
 B_{\ell_1\ell_2\ell_3}&=&\sum_{m_1m_2m_3}
 \threej{\ell_1}{\ell_2}{\ell_3}{m_1}{m_2}{m_3}\,
 \left\langle \tilde{a}_{\ell_1m_1}  \tilde{a}_{\ell_2m_2}  \tilde{a}_{\ell_3m_3} \right\rangle \nonumber\\
 &=& f_{\ell_1\ell_2\ell_3} C_{\ell_2}^{T\phi} C_{\ell_3}^{TT} + \text{5 perm.}.
\end{eqnarray}

\noindent where $C_{\ell}^{TT}$ is the temperature (primordial plus ISW) power spectrum and $C_{\ell}^{T\phi}$ is the cross temperature-lensing angular power spectrum, $C_{\ell}^{T\phi}=\langle\phi^*_{\ell m}a_{\ell m}\rangle$, which depends
on the Weyl potential and its first time-derivative (see e.g.~\cite{spergold}). In deriving~(\ref{angle_averaged_B}) we have implicitly assumed the statistical isotropy of the universe and have averaged the three-point correlation function (in harmonic space) over the orientation of triangles by mean of rotational invariance. Numerical codes evolving perturbations typically work with the reduced bispectrum, defined via
\be\label{red_bisp}
B_{\ell_1\ell_2\ell_3} = \Upsilon_{l_1l_2l_3} b_{\ell_1\ell_2\ell_3}\,.
\ee
 
In Figure \ref{fig:models} we plot different theoretical predictions for $C_\ell^{T\phi}$
and the reduced bispectrum $b_{\ell_1\ell_2\ell_3}$ computed with the publicly available code MGCAMB (\mgcamb). As it can be noticed, the L-ISW bispectrum is clearly sensitive to modifications of gravity and in principle can
be used to put constraints on models of modified gravity. In the next Section we describe the analysis method that we have used to  forecast the latter for Planck-like experiments.

\section {Future constraints from CMB: method} \label{sec:analysis}

We shall estimate the potential of upcoming L-ISW bispectrum measurements from CMB Planck-like experiments
to constrain the modified gravity theories described in the previous section. We perform a likelihood analysis from the spectrum, L-ISW bispectrum and their combination in order to compare the parametrised models of Sec.~\ref{sec:theories} to a fiducial model, chosen to reproduce a $\Lambda$CDM cosmology. We fix the cosmological parameters according to the WMAP 7-year data best fit~\cite{Komatsu:2010fb} and vary only the parameters entering the parametrisations described in Sec.~\ref{sec:theories}. Spanning over the parameter space, we calculate the spectrum and L-ISW bispectrum using MGCAMB and build the likelihoods as described in the following.

Each theoretical model is then compared to the fiducial model with a simple $\chi^2$ function which assumes that the spectrum and bispectrum can be safely described as Gaussian variables~\cite{spergold,giovi}.  
For the standard $C^{TT}_{\ell}$ temperature anisotropy spectrum we have

\begin{equation}
\chi_s^2=\sum_{\ell}^{1000}\left[\frac{C_{\ell}^{TT,th}-C_{\ell}^{TT,fid}}{{\sigma_{\ell}^s}}\right]^2
\end{equation}

\noindent where the uncertainty $\sigma^s$ is given by 

\begin{equation}\label{sigmas}
\sigma_\ell^s=\sqrt{\frac{2}{(2\ell+1)}}C^{TT}_\ell\,,
\end{equation}
\noindent We do not include any covariance noise matrix in Eq.~(\ref{sigmas}), effectively assuming $\sigma^s$ to be cosmic-variance limited up to $\ell = 1000$, which is a good approximation for future Planck-like experiments.

For the bispectrum we have

\begin{equation}
\chi_b^2=\sum_{\ell_1,\ell_2,\ell_3=2}^{l_{\rm max}}\left[\frac{B_{\ell_1\ell_2\ell_3}^{th}-B_{\ell_1\ell_2\ell_3}^{fid}}{{\sigma_{\ell_1\ell_2\ell_3}^b}}\right]^2
\end{equation}
where the sum is over all possible combinations of $\ell_1,\ell_2,\ell_3$ with  $(\ell_1\leq \ell_2 \leq \ell_3)$, $\ell_1+\ell_2+\ell_3$ even and we set $\ell_{\rm max}=1000$, which roughly corresponds to the maximum multipole sensibility for Planck-like experiments, (since at higher multipoles the contamination from foreground point sources 
starts to be dominant). 

\noindent The uncertainty $\sigma_{\ell_1\ell_2\ell_3}^b$ is given by

\begin{equation}
\left({\sigma_{\ell_1\ell_2\ell_3}^b}\right)^2=n_{\ell_1\ell_2\ell_3}C^{TT}_{\ell_1}C^{TT}_{\ell_2}C^{TT}_{\ell_3}\,,
\end{equation}
where $n_{\ell_1\ell_2\ell_3}$ is $6$ for equilateral configurations $(\ell_1=\ell_2=\ell_3)$, $2$ for isoscele ones (with two multipoles equal) and $1$ for the scalene ones (when all the multipoles are different). There is no noise covariance matrix in the $C_{\ell}^{TT}$.

Once the $\chi^2$ functions are computed, we can build the separate likelihoods for the spectrum and bispectrum data respectively: 

\begin{equation}
{\mathcal L}_{s,b}=\exp{\left(-\frac{\chi_{s,b}^2}{2}\right)}
\end{equation}

Neglecting the correlation between spectrum and bispectrum,  we can further combine them in a total likelihood as follows:

\begin{equation}\label{comb_lik}
{\mathcal L}_{c}={\mathcal L}_{s}{\mathcal L}_{b}=\exp{\left(-\frac{\chi_b^2+\chi_s^2}{2}\right)}
\end{equation}

In the calculation of the likelihood from the CMB angular power spectrum we do not include the 
lensing term that is clearly correlated with L-ISW bispectrum. Furthermore, when combining the two likelihoods 
like in (\ref{comb_lik}), we are neglecting correlations between spectrum and bispectrum data that could arise from the large scale
ISW term. As we will see in the next Section this is a good approximation since the bispectrum will constrain modified
gravity parameters with a much stronger significance than spectrum data alone.
When computing the bispectrum we do not include the non-linear Rees-Sciama term, since that would require a modeling of non-linearities in modified gravity.
The exclusion of the RS term is expected to affect our results at most by $\sim 17 \%$, and therefore should not 
 change our conclusions to a significant level. We plan to investigate in a future work the 
non-linear RS term in the framework of modified gravity.

\noindent Finally, since we are modeling the (primordial plus ISW) spectrum as a Gaussian variable, we are effectively neglecting any  inflationary non-Gaussian signal; furthermore, we ignore contributions to the bispectrum from the lensing-Sunyaev-Zel'dovich
correlation. Both signals could anyway be removed exploiting their different angular dependence (see e.g.
\cite{calabrese2}).

For each theoretical model of Sec.~\ref{sec:theories}, while keeping the cosmological parameters fixed to their WMAP 7-year values, we vary the modified gravity parameter(s) (one at a time for the models that have more than one parameter), and compute  the spectrum and the L-ISW reduced bispectrum with  MGCAMB ; we then use Eq.~(\ref{red_bisp}) to compute the L-ISW bispectrum from the reduced one. We also  choose a fiducial model, as discussed in the following, and compute the corresponding spectrum and bispectrum. Once a sufficient number of spectra is calculated, we compute the likelihood profiles and extract the confidence levels on the parameter of interest.

For each parametrisation, we choose a fiducial model based on a 
set of parameters that are, for most of the cases, the parameters that would reduce the cosmology to the
 $\Lambda$CDM one. In the case of the Linder model this is achieved by setting
$\gamma_L=0.555$~\cite{linder2007}.
For $f(R)$ theories,  $B_0=0$ is the value giving $\mu=1=\gamma$ which are the values of these functions in $\Lambda$CDM.
For chameleon theories the choice of the fiducial model is more complicated. Let us start employing the dimensionless parameter $B_0$~(\ref{B0}) in place of the lengthscale $\lambda_1$, so that the parameters for these models become ($B_0$,\,$s$\,,$\beta_1$).
As a matter of fact, we have three free parameters, no strong theoretical reasons to fix two of them and a complete degeneracy among the parameters when trying to reproduce $\Lambda$CDM, {\it i.e.} if we fix either $B_0=0$ or $\beta_1=1$.
We therefore proceed by making a somewhat arbitrary choice on the fiducial model, fixing $\beta_1=1$, $B_0=0.5$, and $s=2$ when studying the constraints on $\beta_1$ and $\beta_1=1.3$, $B_0=0.5$, $s=2$ when studying the 
constraints at varying $s$. 

In the case of Linder's model we evaluate the likelihoods in the range $0.475 \leq \gamma_L \leq 0.635$, at steps of $0.002$ for values near the fiducial one and at steps of $0.01$ for values near the boundaries. In the case of $f(R)$ we explore the likelihood function in the range $0 \leq B_0 \leq 0.7$, varying $B_0$
at steps of $0.1$. In the chameleon case  we use a step of $0.01$ for $\beta_1$ and of $0.2$ for $s$.

\section {Results \& constraints} \label {sec:constraints}

\subsection{Linder model}

\begin {figure}[t]
\includegraphics[width=0.65\linewidth, angle=-90]{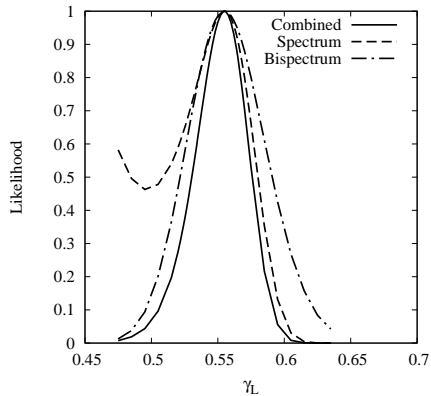}
\caption{Likelihood distribution function for the growth index  $\gamma_L$~(\ref{gammaL}) from 
the analysis of spectrum, bispectrum and combined data.}
\label{fig:liklinder}
\end{figure}

In Table \ref{tab:constraintslin} and Figure \ref{fig:liklinder} 
we report the constraints on  $\gamma_L$ from the
spectrum, the L-ISW bispectrum and the combined analyses.
As we can see the spectrum and bispectrum data are somewhat complementary:
the CMB bispectrum is more powerful in constraining the $\gamma_L$ parameter 
in the region of values lower than those of the fiducial one; 
on the contrary the temperature anisotropy spectrum is more efficient for larger values.
The non-gaussian shape of the likelihood from the temperature spectrum can be 
easily understood by the fact that even in the case of small ISW signal (when $\gamma_L\rightarrow0$) 
the angular spectrum is different from zero and still provides a reasonable fit to the data.
The bispectrum is, on the contrary, not null only if the ISW is different from zero and
it therefore provides a much more reliable way to detect it.

\begin{table}[t]
\begin{center}
\begin{tabular}{|c||c|c|c|c|c|}
\hline 
               &fiducial & {\bf S} & {\bf B}& {\bf C} \\ 
               & & {$95 \%$ c.l.}& {$95 \%$ c.l.}& {$95 \%$ c.l.}\\ 
\hline 
 & & & & \\
\textit{$\gamma_L$} & ${0.555}$&$\begin{array}{c}+0.044\\-\end{array}$ & $\begin{array}{c}+0.060\\-0.056\end{array}$& $\begin{array}{c}+0.034\\-0.042\end{array}$\\ 
& & & &\\
\hline 
\end{tabular} 
\caption{Constraints on $\gamma_L$ of Linder model from the spectrum ({\bf S}) and bispectrum ({\bf B}) and combined ({\bf C}) analyses.}
\label{tab:constraintslin}
\end{center}
\end{table}

As we can see, spectrum data provides solely an upper limit for $\gamma_L$, leaving it practically unconstrained on the lower tail. On the contrary, bispectrum data give a $\sim 5 \%$ error on $\gamma_L$. When spectrum and bispectrum data are 
combined there is a substantial improvement in the measurement.

\subsection{Chameleon models}
\begin{table}[htb]
\begin{center}
\begin{tabular}{|c||c|c|c|c|c|}
\hline 
              &fiducial  & {\bf S} & {\bf B}& {\bf C} \\ 
              &  & {$68 \%$ c.l.}& {$68 \%$ c.l.}& {$68 \%$ c.l.}\\ 
\hline 
 & & & & \\
\textit{$\beta_1$} &$1.00$& $\begin{array}{c}+0.25\\-0.17\end{array}$ & $\begin{array}{c}+0.10\\-0.13\end{array}$& $\begin{array}{c}+0.09\\-0.10\end{array}$\\ 
& & & &\\
\hline 
& & & & \\
\textit{$s$} &$2.00$& $\begin{array}{c}+0.55\\-0.17\end{array}$ & $\begin{array}{c}+0.42\\-0.28\end{array}$& $\begin{array}{c}+0.30\\-0.15\end{array}$\\ 
& & & &\\
\hline 
\end{tabular} 
\caption{Constraints at $1$ standard deviation on the Chameleon models parameters $\beta_1$ and $s$
coming from the analysis of spectrum ({\bf S}), bispectrum ({\bf B}) and combined ({\bf C}) datasets.}
\label{tab:constraintsch}
\end{center}
\end{table}

\begin {figure}[htb]
\includegraphics[width=.65\linewidth, angle=-90]{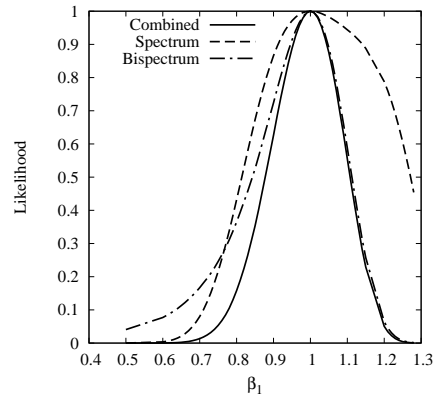}
\includegraphics[width=.65\linewidth, angle=-90]{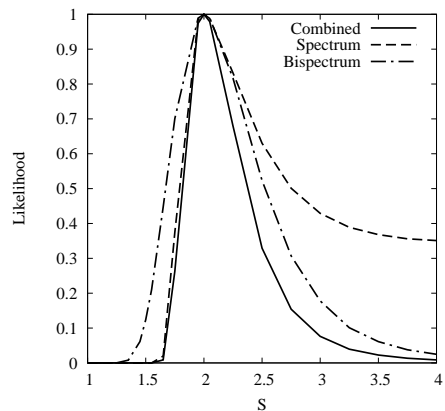}
\caption{Likelihood distribution function for Chameleon models from the spectrum, bispectrum and the combined analyses. The \emph{top panel}  gives the likelihood of the parameter $\beta_1$ when $B_0$ and $s$ are fixed to $B_0=0.5$  $s=2$; the \emph{bottom panel} gives the likelihood for $s$ when $B_0$ and $\beta_1$ are fixed to $B_0=0.5$ and $\beta_1=1.3$.}
\label{fig:likcham}
\end{figure}

The constraints on Chameleon models from the spectrum, L-ISW bispectrum and combined analyses are reported 
in Table \ref{tab:constraintsch} and Figure \ref{fig:likcham}.

As for the Linder model, the two datasets are complementary in constraining the Chameleon parameters.
The simple temperature power spectrum is more powerful in constraining values of 
$\beta_1 \leq 0.75$, {\it i.e.} the lower tail, while the bispectrum data provide stronger 
constraints on the higher tail, where the spectrum data leave the parameter practically
unconstrained.
The same behaviour is seen for the likelihood distribution of the $s$ parameter.
Small values of $s$ ($s<2$) can be better constrained by temperature spectrum data.
However large values of $s$ are left unconstrained from the temperature spectrum and, on the contrary,
are significantly constrained when using the bispectrum. This is related, as we already explained when discussing constraints on $\gamma_L$, to the entity of the ISW signal in the two tails; namely, the spectrum looses constraining power in the parameter range where the ISW is suppressed and tends to zero. 

\subsection{$f(R)$ theories}
\begin {figure}[htb]
\includegraphics[width=0.65\linewidth, angle=-90]{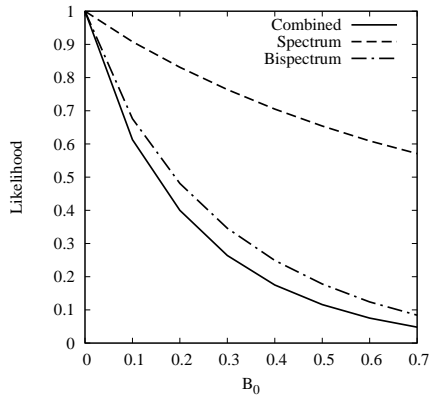}
\caption{Likelihood distribution function for the parameter $B_0$ describing  $f(R)$ theories from 
the analysis of spectrum, bispectrum and combined data.}
\label{fig:likfr}
\end{figure}

In Table \ref{tab:constraintsfr} and Figure \ref{fig:likfr} 
we report the constraints on $f(R)$ models from the
spectrum, L-ISW bispectrum and combined analyses.
In this case the constraints coming from the 
bispectrum are definitely tighter than the ones from the temperature 
spectrum. Once again, this is related to the ISW signal which is suppressed w.r.t. $\Lambda$CDM one for all the values of $B_0$ in the range $0<B_0<3/2$ (becoming null at $B_0=3/2$)~\cite{HuPeiris}. Current constraints from ISW data from CMB-galaxy correlations
are of the order of $B_0<0.4$~\cite{giannantonio2009}. As we show, 
the L-ISW bispectrum can clearly improve CMB constraints on these theories, 
tightening the bounds by a factor of six.

\begin{table}[htb]
\begin{center}
\begin{tabular}{|c||c|c|c|c|c|}
\hline 
              &fiducial  & {\bf S} & {\bf B}& {\bf C} \\ 
             &   & {$68 \%$ c.l.}& {$68 \%$ c.l.}& {$68 \%$ c.l.}\\ 

\hline 
& & & & \\
\textit{$B_0$} &0& $<0.61$ & $<0.14$& $<0.10$\\ 
& & & &\\
\hline 
\end{tabular} 
\caption{Constraints at $1$ standard deviation on $f(R)$ theory parameter $B_0$
coming from the analysis of spectrum ({\bf S}), bispectrum ({\bf B}) and combined ({\bf C}) datasets.}
\label{tab:constraintsfr}
\end{center}
\end{table}

\section {Conclusions} \label {sec:concl}

High accuracy temperature maps of the CMB anisotropy from ongoing and future experiments
will provide an unique opportunity to test non-Gaussianity.
While some of the signal could be primordial, a clear non-Gaussian
signal is expected from the correlation of lensing and the Integrated Sachs-Wolfe effect.
This signal provides a new test of the cosmological scenario
{\it per se}, and could further  be used to test alternatives to the cosmological constant in the context of cosmic acceleration. In this paper we have considered three different parametrisations of modified gravity and
investigated the improvement in constraining their parameters by including the
signal in the bispectrum coming from lensing-ISW correlations.
We have found that in the case when all the cosmological parameters are fixed,
the bispectrum signal will be extremely useful providing a significant
improvement in the constraints on modified theories of gravity.
While the forecasted constraints have been obtained with the assumption of the cosmological concordance model as the fiducial one,  we believe that our results have little dependence
on this choice, since current data accepts only
relatively small deviations from the standard picture.

Finally, while the L-ISW bispectrum signal will be presumably 
detected by the Planck satellite mission at about four standard 
deviations, the CMB lensing signal will be detected at much higher statistical 
significance and could also provide useful constraints on modified 
gravity theories (see e.g.~\cite{cala}). However CMB lensing is
not  directly sensitive to time variations in the gravitational
potentials, which instead enters directly in the L-ISW signal.  Measurements of the ISW signal through correlations of CMB maps
with galaxy surveys already provide interesting constraints on the
models presented here~\cite{giannantonio2009}, however only 
a marginal future improvement in this measurement is expected.
As we have shown, the constraints coming from observations of the L-ISW
bispectrum, being sensitive to both the spatial gradient and the time variation of the Weyl potential,  will be complementary to these other observations, improving CMB bounds on modified theories of gravity.

\subsection*{Acknowledgements}
The work of EDV, AM, VS was supported by the PRIN-INAF grant 'Astronomy probes fundamental physics',
by the Italian Space Agency through the ASI contract Euclid- IC (I/031/10/0). AS is supported by NSF grant No. AST-0708501.

\label{lastpage}
\end{document}